\documentstyle[sprocl,epsfig]{article}

\def\be{\begin{equation}}
\def\ee{\end{equation}}
\def\bea{\begin{eqnarray}}
\def\eea{\end{eqnarray}}

\begin{document}
\title{Vistas in numerical relativity}
\author{M.H.P.M. van Putten}
\address{MIT 2-378, Cambridge, MA 02139, USA\\
E-mail: mvp@schauder.mit.edu}

\maketitle\abstracts{ 
Upcoming gravitational wave-experiments promise a window for discovering
new physics in astronomy. Detection sensitivity of the broadband 
laser interferometric detectors LIGO/VIRGO may be enhanced by
matched filtering with accurate wave-form templates. Where analytic methods
break down, we have to resort to numerical relativity, often in Hamiltonian
or various hyperbolic formulations.  
Well-posed numerical relativity requires consistency with the elliptic constraints
of energy and momentum conservation. We explore this using a choice of gauge 
in the future and a dynamical gauge in the past. Applied to a polarized Gowdy 
wave, this enables solving {\em all} ten vacuum Einstein equations.
Evolution of the Schwarzschild metric in 3+1 and, more generally, 
sufficient conditions for well-posed numerical 
relativity continue to be open challenges.
}

\section{Introduction}

The Laser Interferometric Gravitational-wave Observatories LIGO/VIRGO
\cite{abr92,bra91} is a broad band detector, targeting 
gravitational radiation from compact sources of a few solar masses
\cite{tho95}. Notable 
sources are binary coalescence of neutron stars and black holes, as well as 
emissions from black hole-torus systems as recently proposed \cite{mvp02}. 
Their gravitational wave emissions 
provide a record of these strongly interacting catastrophic events and could 
contain most of the total energy released. What is the first source 
that LIGO/VIRGO may detect?
Our knowledge of the gravitational wave-forms may be the determining factor
in answering this question. For this reason,
numerical simulations of general relativity (numerical relativity \cite{leh01})
are receiving much attention in efforts towards matched filtering 
in searches for binary coalescence involving neutron stars and black holes 
\cite{bru00}.

\subsection{Some astrophysical problems for numerical relativity}

There exists a broad spectrum of candidate sources of gravitational waves.
Many sources have been proposed, for instance the coalescence of binaries of
neutron stars and black holes \cite{phi91,nar91,bel02}, supernovae (see \cite{fer98}),
and rapidly spinning neutron stars \cite{and98,lin01,arr02}.

The gravitational wave-forms produced by neutron star-neutron star coalescence
in the inspiral phase are well-understood with post-Newtonian expansion technique
(see \cite{bla02} for a recent review). Black hole-black hole coalescence is very
promising because of the expected larger amplitude signal. However, their event
rate is highly uncertain \cite{bel02}. 
Their wave-forms in the merger phase, believed to be
relevant for LIGO/VIRGO detection, is not well understood. This is left 
as a challenge for numerical relativity (see, e.g., \cite{abr98,coo98,gom98}). 

A new model was recently proposed for gravitational radiation from at torus around
a black hole \cite{mvp01b,mvp01d,mvp02}. One particular feature is that it is 
associated with long gamma-ray bursts, whose intrinsic durations are about 20s on
average. These energetic events are observed at a rate 
of about 2 per day and, corrected for beaming, at a rate of about 1 per year within
a distance of 100Mpc.
An important problem is the nature of the torus' mass-quadrupole 
moment, which may be determined by numerical relativity. 

\subsection{Numerical relativity: {integration on thin ice}}

 Long-time 
 integrations for the purpose of matched filtering requires accurate 
 conservation of the elliptic constraints representing energy and momentum 
 conservation. This is a natural requirement, which becomes apparent in
 numerical experiments \cite{kid01}.
 
 The intrinsic hyperbolic structure of wave-motion permits well-posed initial value problems,
 as follows from the energy method. In the continuum limit, conservation of energy
 and momentum is exact and drops out of the ``energy balance sheet''. Consequently,
 well-posedness for general relativity reduces to the anticipated results for its 
 hyperbolic structure.

 The nonlinear nature of the Einstein equations tends to introduce numerically a
 departure from exact conservation of the elliptic constraints. 
 {\em And the initial value problem for non-hyperbolic equations is often ill-posed, 
 as for instance the Laplace and backward heat equation.}
 The familiar Hadamard counter example to well-posedness is given by the solution
\begin{eqnarray}
u_{n}=\frac{1}{n}\sin n x\sinh ny
\end{eqnarray}
 of Laplace's equation on the upper half-plane:
 \begin{eqnarray}
 u_{xx}+u_{yy}=0~~~(-\infty<x<\infty,~~y\ge0)
 \end{eqnarray}
 subject to the Cauchy data
 \begin{eqnarray}
 u(x,0)=0,~~u_y(x,0)=\frac{1}{n}\sin nx.
 \end{eqnarray}
 The solution blows up in the face of large $n$, even though the initial data 
 approach zero in the norm of continuously differentiable functions. Numerically, 
 this tends to result in ill-posedness: rapidly growing errors, regardless of the 
 accuracy of initial conditions. The backward heat 
 equation serves to illustrate similar ill-posedness in the presence of a 
 first-order time-derivative.
 
 The observed instabilities associated with constraint violations \cite{kid01} 
 suggests a link between variables with time-derivatives in 
 the energy-momentum constraints and ill-posedness.
 Well-posed numerical relativity requires these constraints to be evolved in
 a consistent manner. A key test is the `dynamical' evolution of a 
 Schwarzschild black hole, where the dynamics derives from a singularity
 avoiding foliation of spacetime (see, e.g., \cite{kid01}). 
 In the absence of a covariant separation of the
 hyperbolic-elliptic structure of general relativity, we shall in this lecture
 discuss a recent proposal for an advanced-retarded evolution of the Einstein 
 equations for exact preservation of the constraints under dynamical evolution.

\section{The Einstein equations, spacetime foliation and conservation laws}

   The Einstein equations describe the structure of a curved spacetime 
   manifold $M$ with four-covariant metric $g_{ab}$ in response to a 
   stress-energy tensor $T_{ab}$. They are 
\begin{eqnarray}
R_{ab}-\frac{1}{2}g_{ab}R = 8\pi T_{ab}
\label{EQN_EE}
\end{eqnarray}
as an equation for the Ricci tensor $R_{bd}=R^c_{\cdot bcd}$ and its 
scalar curvature $R=R^c_{\cdot c}$. These are expressions in terms of 
the Riemann tensor $R^{a}_{\cdot bcd}$. The left hand-side is commonly 
denoted by the Einstein tensor $G_{ab}$, with the property that 
$\nabla^aG_{ab}\equiv0$ (the Bianchi identity) is consistent with
energy-momentum conservation $\nabla_aT^{ab}=0$. We are at liberty to 
include a cosmological constant term $-\Lambda g_{ab}$ on the right 
hand-side of (\ref{EQN_EE}). The equations of motion (\ref{EQN_EE})  
derive from the Hilbert action
\begin{eqnarray}
{\cal S}=\int_M \sqrt{-g} R d^4x.
\label{EQN_LA}
\end{eqnarray}
Translation invariance is a symmetry in this action (\ref{EQN_LA}).
By Noether's theorem, this leads to four conservation laws of energy and 
momentum. The associated gauge group is the Lorentz group SO(3,1,$R$) of 
boosts and rotations. These conservation laws become explicit on spacelike 
hypersurfaces $\Sigma_t$: $t=$const., where $t$ denotes a timelike coordinate. 
Combined, the surfaces $\Sigma_t$ provide a foliation of spacetime.

\subsection{Spacetime foliation in spacelike hypersurfaces $\Sigma_t$}

Hypersurfaces $\Sigma_t$ of constant time come with two vectors:
 \begin{eqnarray}
 {\cal N}_a=g_{ta},~~~n_a=\partial_a t/\sqrt{-\partial_at\partial^at},
 \end{eqnarray}
 where $n_a$ denotes the unit normal ($n^2=-1)$ to $\Sigma_t$. 
 (The vector ${\cal N}_a$ is commonly denoted by $t_a$ \cite{wal84}.)
 Generally, the covariant vectors ${\cal N}_a$ and $n_a$ are independent.
 Marching from one hypersurface to the next brings along a variation $d t$, 
 along with the covariant displacement
 \begin{eqnarray}
 d s_{a}={\cal N}_a d t.
\label{EQN_SA}
 \end{eqnarray}
 The displacement $ds_{(a)}$ expresses ${\cal N}_a$ as a ``flow of time." It
 can be expressed in terms of orthogonal projections on $n_a$ and $\Sigma_t$,
 in terms of the lapse function $N$ and shift functions $N_a$,
  \begin{eqnarray}
{\cal N}^a=Nn^a+N^a.
\label{EQN_TIME}
 \end{eqnarray}
 Here $N=-{\cal N}_an^a$ and $N_a=h_{a}^b{\cal N}_b$, expressed in the metric
\begin{eqnarray}
h_{ab}=g_{ab}+n_an_b
\label{EQN_GH}
\end{eqnarray}
as the orthogonal projection of $g_{ab}$ onto $\Sigma_t$. 
Note that $ds^2={\cal N}^2dt^2=g_{tt}dt^2$ as the square of (\ref{EQN_SA}),
so that $g_{tt}=-N^2+N_cN^c$. With $n_a=(n_t,0,0,0)$, it follows that
\begin{eqnarray}
g_{ab}=
\left(\begin{array}{cc}
N^cN_c-N^2 & N_j\\
N_i & h_{ij}
\end{array}\right),
\label{EQN_GN}
\end{eqnarray}
where $i,j$ refer to the spatial coordinates $x^i$ of $(t,x^i)$.
The lapse function satisfies $\sqrt{-g}=N\sqrt{h}$. 
The four degrees of freedom in the five functions $(N,N_a)$ are algebraically 
equivalent to ${\cal N}_a$. An equivalent expression for the line-element,
in so-called 3+1 form \cite{tho86}, is 
\begin{eqnarray}
ds^2=-\alpha^2 dt^2 + \gamma_{ij}(dx^i+\beta^i)(dx^j+\beta^j),
\label{EQN_TPO}
\end{eqnarray}
where $\alpha=N$ is referred to as the redshift factor and 
$\gamma_{ij}\beta^j=g_{it}$.

\subsection{Conservation of energy and momentum}

Coordinate invariance introduces a certain degeneracy in the Einstein equations.
There are no second-order time-derivatives of $g_{ab}$ in the components 
$G_{nb}=G_{ab}n^a$ of the Einstein tensor $G_{ab}=R_{ab}-(1/2)g_{ab}R$.
Consequently, the expression $G_{nb}$ forms entirely out of Cauchy data
on $\Sigma_t$ (data and their first time-derivatives). 
The embedding of $\Sigma_t$ in four-dimensional spacetime is expressed in 
terms of the symmetric extrinsic curvature tensor $K_{ab}$. If $\tilde{n}_b$ 
denotes a unit tangent to a geodesic orthogonal to $\Sigma_t$, then
\begin{eqnarray}
K_{ab}=\nabla_a \tilde{n}_b=\frac{1}{2}{\cal L}_n h_{ab}.
\label{EQN_K}
\end{eqnarray}
Thus, $K_{ab}$ represents a time-like derivative of the metric in $\Sigma_t$,
which is a velocity of $h_{ab}$. 
We use here the sign convention in \cite{wal84}; $K_{ab}$ with opposite sign
is commonly used in numerical relativity. We then have
\begin{eqnarray}
\begin{array}{rl}
D_bK^b_{\cdot a}-D_aK & =8\pi T_{an},\\
\mbox{}^{(3)}R+K^2-K_{ab}K^{ab} & =16\pi T_{nn}
\end{array}
\label{EQN_CL}
\end{eqnarray}
a consequence of the projection $^{(3)}R^{a}_{\cdot bcd}$ of the 
four-dimensional Riemann tensor $R^{a}_{\cdot bcd}$ on $\Sigma_t$ 
(see \cite{yau81,wal84,mvp96} for detailed calculations). Here,
$K=K_c^c$ denotes the trace of $K$. These expressions (\ref{EQN_CL})
are, respectively, the conservation laws of linear momentum and
energy. These equations are elliptic in the spatial coordinates
internal to $\Sigma_t$.

\section{Two marching methods for hyperbolic formulations}

 A practical frame-work for numerical relativity consists of marching
 data from one space-like hypersurface $\Sigma_t$: $t=$const. to the next. 
 The hypersurface $\Sigma_t$ is generally dynamical.
 A slicing of spacetime either comes {\em before} or {\em after} a
 choice of dynamical variables. Chosen before, the variables live
 in $\Sigma_t$ as projections of the underlying four-covariant metric. 
 Chosen after, one continues to work with four-covariant metric
 parametrized
 over the hypersurfaces $\Sigma_t$. There is no dictum for the order of these 
 choices, but they do give manifestly different formulations.

\subsection{Slice first: the Hamiltonian approach}

In the Hamiltonian approach, we consider first a choice of foliation of spacetime in 
spacelike hypersurface $\Sigma_t$ of constant coordinate time $t$. Their
dynamics is described by the projected metric $h_{ab}$ with canonical
momentum $\pi_{ab}$, satisfying the Hamiltonian equations
\begin{eqnarray}
\dot{h}_{ab} = \frac{\delta {H}}{\delta \pi^{ab}},~~~
\dot{\pi}^{ab} = -\frac{\delta {H}}{\delta h_{ab}},
\end{eqnarray}
where the dot denotes the Lie derivative ${\cal L}_t$ with
respect to the vector field $t^a={\cal N}^a$ of the flow of time 
(\ref{EQN_TIME}). This Lie derivative reduces to differentiation with respect 
to $t$ in our coordinate system $(t,x^i)$.

The Hamiltonian equations derive from the Hilbert action (\ref{EQN_LA}).
An excellent presentation is given in Appendix E in \cite{wal84}, which is 
briefly summarized here.

The Lagrangian density ${\cal L}=R\sqrt{-g}$ can be expressed as 
a sum of the three-curvature $^{(3)}R$ of $\Sigma_t$ and a quadratic form 
of the extrinsic curvature tensor $K_{ab}=\nabla_an_b$, given by
${\cal L}=\sqrt{h}N[^{(3)}R+K_{ab}K^{ab}-K^2].$
We have \cite{wal84} $K_{ab}=L_nh_{ab}/2=
[\dot{h}_{ab}-D_aN_b-D_bN_a]/(2N),$
where $D_a=h^b_a\nabla_a$ denotes the derivative internal to $\Sigma_t$.
It follows that
\begin{eqnarray}
\pi^{ab}=\frac{\partial{\cal L}}{\partial\dot{h}}=
         \sqrt{h}(K^{ab}-Kh^{ab}).
\end{eqnarray}

Coordinate invariance of the Einstein equations leaves $g_{ta}$ and, hence, 
$(N,N_a)$ freely specifyable.
Tracing back, we indeed find no first-order time derivatives of $g_{tb}$ 
in the Lagrangian density ${\cal L}$, whereby the associated canonical 
momenta vanish. The Hamiltonian density associated with the dynamical degrees of
freedom reads therefore ${\cal H}=\pi^{ab}\dot{h}_{ab}-{\cal L}.$
The variational derivative of $H=\int_{\Sigma_t}{\cal H}\sqrt{h}d^3x$ with
respect to these gauge functions obtains the conservation laws of energy
and momentum (\ref{EQN_CL}). The variational derivative with respect
to the dynamical variables $(h_{ab},\pi^{ab})$ obtains the ADM
formulation \cite{arn62,wal84}
\begin{eqnarray}
\dot{h}_{ab}
-2D_{(a}N_{b)}=
2h^{-1/2}N(\pi_{ab}-\frac{1}{2}h_{ab}\pi)
\label{EQN_ADM1}
\end{eqnarray}
and
\begin{eqnarray}
\begin{array}{rll}
\dot{\pi}^{ab}
-2\pi^{c(a}D_cN^{b)}
&=&-Nh^{1/2}(^{(3)}R-\frac{1}{2}^{(3)}Rh^{ab})\\
  &&+h^{1/2}
                 (D^aD^bN-h^{ab}D^cD_cN)
 \\&&+\frac{1}{2}Nh^{-1/2}h^{ab}(\pi:\pi-\frac{1}{2}\pi^2)
   \\&&-2Nh^{-1/2}(\pi^{ac}\pi_c^b-\frac{1}{2}\pi\pi^{ab})\\
&&+h^{1/2}D_c(h^{-1/2}N^c\pi^{ab}).
\end{array}
\label{EQN_ADM2}
\end{eqnarray}

In numerical relativity, these Hamiltonian evolution equations are often considered in
terms of the non-canonical pair $(\gamma_{ij},K_{ij})$, with
$\gamma_{ij}$ as in (\ref{EQN_TPO}) and $K_{ij}$, where $i,j$ refer to the spatial
components in $(t,x^i)$. Thus, (\ref{EQN_ADM1}) and (\ref{EQN_ADM2}) become,
using the vacuum case of (\ref{EQN_CL}),
\begin{eqnarray}
\begin{array}{ll}
(\partial t-{\cal L}_\beta)\gamma_{ij}&=-2\alpha K_{ij},\\
(\partial_t-{\cal L}_\beta)K_{ij}&=-D_iD_j\alpha+\alpha[^{(3)}R_{ij}-2K_{ij}^2+KK_{ij}],
\end{array}
\label{EQN_NC}
\end{eqnarray}
where we use the definition of $K_{ij}$ with opposite sign 
of (\ref{EQN_K}), following the convention in this context.

\subsection{Hyperbolic formulations in the Hamiltonian approach}

Hyperbolic systems of equations in the
non-canonical variables
$(\gamma_{ij},K_{ij})$, or closely related variables,
have been derived in various forms by several groups, notably
\cite{fri94,cho95,abr95,bon95,fri96,and97,alc99,fri99,and99,her00}; see  
\cite{reu98} for a comprehensive review.
This approach typically comprises constraints on the lapse function
\cite{abr95}. For a recent comparison study between formulations in
Hamiltonian variables and related hyperbolic formulations, see
\cite{bar01,bon02}. 
Different formulations display various 
degrees of numerical stability \cite{sch98,bau99}, 
which appears not to depend significantly on the degree of hyperbolicity
\cite{shi00,yon00}.

\subsection{Dynamical conservation of constraints}

The evolution equations for general relativity may formally be modified,
such that the energy and momentum constraints
become a stable manifold of physical solutions. This has recently been
considered in a linearized treatment \cite{bro99} and in
Ashtekar's formulation \cite{shi00,yon00,shi01}. This holds some promise in
providing a unified treatment of the dynamical and the elliptic
parts of general relativity. 
Numerical results on accuracy and stability are inconclusive at present
\cite{sie01}.

\subsection{Slice last: the four-covariant approach}

General relativity can be written as nonlinear wave equations 
for the Riemann-Cartan connections in the tetrad formulation.
This builds on Pirani's arguments concerning the role of
the Riemann tensor in gravitational waves \cite{pir57} and on Yang-Mills
formulations of general relativity, following Utiyama \cite{uti57}
and developed by Ashtekar and co-workers \cite{ash86,ash91,got92}.
Starting point is a divergence equation for the Riemann tensor with 
an anti-symmetric derivative of the stress-energy tensor as a source-term.

The interwoveness of wave motion and causal structure distinguishes
gravity from other field theories. This becomes
apparent in nonlinear wave equations for the connections on the curved
spacetime manifold side-by-side with equations of structure for the
evolution of the metric in the tangent bundle. 

The tetrad approach \cite{mvp96,est97} bears some relation to but is 
different from Ashtekar's propram on nonperturbative quantum gravity.
The original Ashtekar variables are SU(2,C) soldering
forms and an associated complex connection in which the constraint
equations become polynomial. The Riemann-Cartan variable is a real
SO(3,1,R) connection. In Ashtekar's variables, a real spacetime
is recovered from the complex one by reality constraints. See
Barbero for a translation of Ashtekar's approach into SO(3,R) phase
space with real connections \cite{bar94,bar95}. 
The main innovation in \cite{mvp96} is the
incorporation of the Lorentz gauge condition (\ref{EQN_LG}) 
which obtains new hyperbolic evolution equations in four-covariant form
(below).


Following Pirani, we take the view that gravitational wave-motion is
contained in the Riemann tensor, $R_{abcd}$. It satisfies the Bianchi
identity
\begin{eqnarray}
3\nabla_{[e}R_{ab]cd}=\nabla_eR_{abcd}+\nabla_aR_{bacd}+\nabla_{b}R_{eacd}=0.
\label{EQN_YM0}
\end{eqnarray}
This gives rise to the homogeneous divergence equation
$\nabla^a*R_{abcd}=0,$
where $*R_{abcd}=(1/2)\epsilon_{ab}^{\cdot\cdot ef}R_{efcd}$ denotes
its dual. Upon interaction with matter in accord with the Einstein
equations, the Ricci tensor satisfies
$R_{ab}=8\pi[T_{ab}-\frac{1}{2}g_{ab}T].$
The Bianchi identity above also gives 
$\nabla^dR_{abcd}=2\nabla_{[b}R_{a]c}$,
which obtains the inhomogeneous divergence equation
\begin{eqnarray}
\nabla^aR_{abcd}=16\pi(\nabla_{[c}T_{d]b}-\frac{1}{2}g_{b[d}\nabla_{c]}T).
\label{EQN_DR}
\end{eqnarray}
The quantity on the right hand-side shall be referred to as $16\pi\tau_{bcd}$.
This term is divergence free, $\nabla^b\tau_{bcd}\equiv0$, on account of the
conservation law $\nabla^aT_{ab}=0$ and consistent with
divergence-free condition $\nabla^b\nabla^aR_{abcd}=0$ on the left hand-side
(\ref{EQN_DR}) (by anti-symmetry of
the Riemann tensor in its first two indices).


Introduce the Riemann-Cartan connections
$\omega_{a\mu\nu}=(e_\mu)^c\nabla_a(e_\nu)_c$
associated with a tetrad $\{(e_\mu)^b\}_{\mu=1}^4$. Then the above-mentioned
homogeneous and inhomogeneous divergence equations take the form
\begin{eqnarray}
\hat{\nabla}^a*R_{ab\mu\nu}=0, ~~~\hat{\nabla}^aR_{ab\mu\nu}=16\tau_{b\mu\nu},
\label{EQN_YM3}
\end{eqnarray}
where the $\omega_{a\mu\nu}$ define a gauge-covariant
derivative in accord with the Yang-Mills construction
$\hat{\nabla}_a=\nabla_a+[\omega_a,\cdot].$
The first of (\ref{EQN_YM3}) gives rise to the representation
$R_{ab\mu\nu}=\nabla_a\omega_{b\mu\nu}-\nabla_{b}\omega_{a\mu\nu}
             +[\omega_a,\omega_b]_{\mu\nu}.$
The gauge covariant derivative satisfies the identity
$\hat{\nabla}_a(e_\mu)^b \equiv0$, which implies the equations of structure
$\partial_{[a}(e_\mu)_{b])}=(e^\nu)_{[b}\omega_{a]\nu\mu}$ -- 
leaving $\partial_t(e_\mu)_t$ undefined. 
Next, define $\xi^b=(\partial_t)^b$,
and introduce the {\em tetrad lapse functions}
\begin{eqnarray}
N_\mu=(e_\mu)_a\xi^a
\end{eqnarray}
as freely specifiable functions. Thus, the equations of structure become
a system of first-order ordinary differential equations
\begin{eqnarray}
\partial_t(e_\mu)_b+\omega_{t\mu}^{\cdot\cdot\nu}(e_\nu)_b=
  \partial_bN_\mu+\omega_{b\mu}^{\cdot\cdot\nu}N_\nu.
\label{EQN_YM5}
\end{eqnarray}
The tetrad lapse functions are algebraically equivalent to the
familiar Hamiltonian lapse, $N$, and shift functions, $N_a$, through
(\ref{EQN_GN}): 
\begin{eqnarray}
g_{at}=N_\alpha(e^\alpha)_a=(N_qN^q-N^2,N_p).
\end{eqnarray}
The term $\omega_{b\mu\nu}N^\nu$ on the right hand-side of (\ref{EQN_YM5})
shows that the tetrad lapse functions introduce different transformations
on each of the legs; the term $\omega_{t\mu}^{\cdot \cdot \nu}
(e_\nu)_b$ on the left hand-side introduces a transformation with applies
to all four legs simultaneously. It follows that it is the infinitesimal
Lorentz transformations $\omega_{t\mu}^{\cdot\cdot}$ which provide the
internal gauge transformations.

\subsection{Hyperbolic equations in the four-covariant approach}

An important aspect here is internal gauge fixing on the
Lorentz group SO(3,1,$R$) associated with the choice of tetrad. 
To fix gauge, we propose using the Lorentz gauge \cite{mvp96}
\begin{eqnarray}
c_{\mu\nu}:=\nabla^a\omega_{a\mu\nu}=0.
\label{EQN_LG}
\end{eqnarray}
This fixes unique evolution equations for the internal gauge,
resulting in nonlinear wave equations for the connections
$\omega_{a\mu\nu}$.
These complement the
equations of structure for the evolution of the tetrad legs,
and together form a complete system of evolution equations. 
The Lorentz gauge (\ref{EQN_LG}) 
defines a choice of {\em acceleration} of the tetrad legs,
through the infinitesimal Lorentz transformations $\omega_{t\mu\nu}$
mentioned above.
In a different context of compact gauge groups and a metric
with Euclidean signature, its geometric significance has been
interpreted by Lewandowski et al. (1983) \cite{lew83}. These six
constraints  (\ref{EQN_LG}) can be given a hyperbolic implementation
by application of the divergence technique \cite{mvp91,mvp96b}
\begin{eqnarray}
\hat{\nabla}^a\{R_{ab\mu\nu}+g_{ab}c_{\mu\nu}\}=16\pi\tau_{b\mu\nu},
\label{EQN_YM7}
\end{eqnarray} 
which preserves $c_{\mu\nu}=0$ is preserved 
in the future domain of dependence of the support of physical initial data. 
By explicit calculation, we have
\begin{eqnarray}
\hat{\Box}\omega_{a\mu\nu}-R_a^c\omega_{c\mu\nu}
-[\omega^c,\nabla_a\omega_c]_{\mu\nu}=16\pi\tau_{a\mu\nu},
\end{eqnarray}
where $\hat{\Box}$ denotes the Yang-Mills wave operator $\hat{\nabla}^2$.
The Ricci tensor on the left hand-side may be understood in terms of 
$T_{ab}$ using the Einstein equations. 
The above provides the following covariant separation \cite{mvp96}.\\
\mbox{}\\
{\bf Theorem 1.} {\em Gravitational waves propagate on a curved
spacetime manifold by nonlinear wave equations in a Lorentz gauge
on the Riemann-Cartan connections. In response to this wave motion, 
the causal structure of the manifold evolves in the tangent bundle 
by the equations of structure. The Hamiltonian lapse and shift functions
find their algebraic counterparts in the tetrad lapse functions $N_\mu$.}

\mbox{}\\
We remark that away from the matter source, the vacuum equations read
\begin{eqnarray}
\hat{\Box}\omega_{a\mu\nu}-[\omega^c,\nabla_a\omega_c]_{\mu\nu}=0.
\end{eqnarray}
This vacuum case has been considered numerically in one-dimensional
Gowdy-tests \cite{mvp97} by using the underlying first-order system for the
components of the Riemann tensor (\ref{EQN_YM7}).

\section{Hyperbolic Einstein-MHD equations}

Relativistic hydrodynamics and magnetohydrodynamics has received
considerable attention in the simulations of astrophysical jets
\cite{mar99,gom01}. Recently, these efforts result in
simulations of astrophysical jets around black holes \cite{mei01} with
extensions to flows around rotating black holes for a few dynamical 
time-scales \cite{koi02}. The latter shows a transition of accretion
disk outflows towards a state of differential rotation in the vicinity 
of the black hole. 

GRBs from rotating black holes are associated with a compact torus 
or disk, representing binary black hole-neutron 
star coalescence \cite{pac91}, failed-supernovae \cite{woo93} or 
hypernovae \cite{pac98,bro00}. The torus or disk may well be magnetized
with the remnant field of the progenitor star, i.e., the neutron star in the
coalescence scenario or young massive star in the failed-supernova or
hypernova scenario. The suggests simulating the creation of gravitational
waves by high-density matter around a black hole in the approximation of
ideal magnetohydrodynamics.

A hyperbolic formulation of ideal magnetohydrodynamics, used in the
simulation of relativistic jets \cite{mvp96}, is
\begin{eqnarray}
\begin{array}{rl}
\nabla_a T^{ab}&=0,\\
\nabla_a(u^{[a}h^{b]}+g^{ab}c)&=0,\\
\nabla_a(ru^a)&=0,
\end{array}
\label{EQN_MHD}
\end{eqnarray}
expressing conservation of energy-momentum, Faraday's equations in
divergence form which conserves the constraint $c=u^ch_c=0$ \cite{mvp91}, 
and conservation of baryon number.
This is a single fluid description of an ideal, inviscid fluid
with stress-energy tensor \cite{lic67}
\begin{eqnarray}
T_{ab}=(r+\gamma P/(\gamma-1) + h^2)u^au^b +(P+h^2/2)g^{ab} -h^ah^b,
\end{eqnarray}
where $u^a$ denotes the velocity four-vector, $r$ the comoving
rest mass density and $P$ the pressure with polytropic equation of state
$P=Kr^\gamma$ and polytropic index $\gamma$.

Perhaps Theorem 1 and (\ref{EQN_MHD}) may serve as a starting point
for hyperbolic Einstein-MHD equations for the purpose of numerical
simulations.

 \section{Past and future gauge in numerical relativity}

 The Einstein equations pose six equations for dynamical evolution plus
 the four constraints of energy and momentum conservation. The latter
 involve the normal vector $n^a$ to the surfaces of foliation, i.e.:
 the projection operator onto these surfaces. The six dynamical degrees of
 freedom pertain to variables subject to second-order time-derivatives,
 while gauge-variables are subject to constraints on their first-order
 time-derivatives. In a discrete setting, the first live on three and the
 second on two hypersurfaces of constant time. This suggests to consider
 the ten degrees of freedom involved to be the six dynamical degrees
 of freedom supplemented with four gauge-variables {\em from the past}.
 The gain is exact conservation of energy-momentum, traded off against 
 exact projections in the past. 

 Non-exact projections naturally permit an uncertainty between
 the three-metric and its canonical momentum within the underlying context
 of a four-covariant theory, i.e.: also in regards to the association with 
 the hypersurface at hand. In the covariant approach of \cite{mvp96}, this 
 would thus reflect an uncertainty in the tetrad elements, which define the 
 projection, and their connections.
 This points towards a potential connection to quantum gravity. Indeed, 
 soon after this work was proposed \cite{mvp01_a}, the author learned of
 a very interesting independent discussion on the problem of consistent 
 discretizations in this context \cite{gam01}.

 \subsection{A discretized initial value problem}
 
 We illustrate our this approach on the vacuum Einstein
 equations
 \begin{eqnarray}
 R_{ab}=0.
 \label{EQN_A}
 \end{eqnarray}
 The Ricci tensor $R_{ab}$ is a second-order expression in the 
 metric $g_{ab}$, whereby (\ref{EQN_A}) defines a relationship between
 metric data $(g^{n-1}_{ab},g^n_{ab},g^{n+1}_{ab})$ on a triple of 
 time-slices $t_{n-1}<t_n<t_{n+1}$:
\begin{eqnarray}
 R_{ab}\left(g_{ab}^{n+1},g_{ab}^n,g^{n-1}_{ab} \right) = 0.
\end{eqnarray}
 Here, $R_{bd}=R^a_{\hskip0.1in bcd}$ derived from the Riemann tensor
 \begin{eqnarray}
 R^a_{\hskip0.1in bcd}=\partial_d\Gamma^a_{bc}-\partial_c\Gamma^a_{bd}
		    +\Gamma^e_{bc}\Gamma^a_{ed}
					-\Gamma^e_{bd}\Gamma^a_{ec}.
 \label{EQN_R}
 \end{eqnarray}
 This expression (\ref{EQN_R}) can be discretized by finite differencing
 on a triple of time-slices with preservation of the quasi-linear 
 second-order structure of $R_{ab}$.

 Algebraic gauge-fixing takes the form of specifying the components
 $N_a=g_{ta}$ in coordinates $\{x^a\}_{a=1}^{4}$ with $t=x^1$ time-like.
 A gauge-choice on a triple of time-slices amounts to a 
 choice of $(N_a^{n-1},N_a^n,N_a^{n+1})$. This gauge-choice 
 in the metric arises explicitly in the Hamiltonian constraints of 
 energy-momentum conservation. The components $h_{ij}=g_{ij}$, where 
 $i,j=2,3,4$ refer to projections of the metric into the time-slice
 $t=$const., which describe the dynamical part of the metric. The
 combination $(h_{ij},N_a)$ reflects 
 the mixed hyperbolic-elliptic structure in numerical relativity and
 (\ref{EQN_A}) represents ten evolution equations in these variables
 on a triple of time-slices. 
 
 In algebraic gauge-fixing, we prescribe $N_a^{n+1}$ as a future gauge 
 in computing $h^{n+1}_{ij}$ on a future hypersurface $t=t_{n+1}$ from 
 data at present and past hypersurfaces $t=t_{n-1}$ and $t=t_{n}$.  
 Re-introducing $N_a^{n-1}$ as dynamical gauge
 in the past gives closure, leaving $h_{ij}^{n-1}$ fixed. This combination
 of ten degrees of freedom
 defines an advanced hyperbolic-retarded elliptic evolution of the metric.
 The paritioning of the metric in past and future variables as
 \begin{eqnarray}
 g_{ab}=(h_{ij}^{n+1},N_a^{n-1})=\left(
 \begin{array}{cccc}
 N_1^{n-1} & N_2^{n-1} & N_3^{n-1} & N_4^{n-1}\\
 N_2^{n-1} & h_{xx}^{n+1} & h_{xy}^{n+1} & h_{xz}^{n+1}\\
 N_3^{n-1} & h_{xy}^{n+1} & h_{yy}^{n+1} & h_{yz}^{n+1}\\
 N_4^{n-1} & h_{xz}^{n+1} & h_{zy}^{n+1} & h_{zz}^{n+1}
 \end{array}
 \right)
 \label{EQN_PG}
 \end{eqnarray}
 thus obtains ten dynamical variables in the ten equations
 \begin{eqnarray}
 R_{ab}(h_{ij}^{n+1},N_a^{n-1},\cdots)=0~~\mbox{at}~~t=t_n.
 \label{EQN_B}
 \end{eqnarray}
 The dots refer to the remaining data
 $(h_{ij}^{n-1},h_{ij}^n,N_a^n,N_a^{n+1})$, which are kept fixed 
 while solving for $(h_{ij}^{n+1},N_a^{n-1})$.
 Thus, (\ref{EQN_B}) which takes into account {\em all} ten Einstein
 equations with no reduction of variables. Time-stepping by (\ref{EQN_B}) 
 {\em evolves the metric into the future with dynamical gauge in the
 past}, in an effort to satisfy energy-momentum conservation within the
 definition of the discretized Ricci tensor.
Because (\ref{EQN_B}) comprises derivatives of $N_a$ only to first-order 
in time, numerically though the data $N^{n+1}_a$ and $N^{n-1}_a$,
we anticipate that the evolution of $N_a$ is of first-order 
in the $t-$discretization $\Delta t$. This introduces
non-exactness in $h_{ij}^{n-1}$ as projections of $g_{ab}$ on 
$t-\Delta t$ to within the same order of accuracy. It may result in a 
first-order drift in the $t-$labeling of the hypersurfaces -- 
permitted by coordinate invariance.
 
\subsection{A polarized Gowdy wave}

 The presented approach can be illustrated on a polarized Gowdy wave.
 Gowdy cosmologies have 
 compact space-like hypersurfaces with two Killing vectors 
 $\partial_\sigma$ and $\partial_\delta$. With cyclic boundary conditions, 
 the space-like hypersurfaces are homeomorphic to the three-torus as a 
 model universe collapsing into a singularity. The associated line-element 
 is (see, e.g., \cite{ber93})
 \begin{eqnarray}
 ds^2=e^{(\tau-\lambda)/2}\left(-e^{-2\tau}d\tau^2+d\theta^2\right)
      +d\Sigma^2,
\label{EQN_G1}
 \end{eqnarray}
 where $\lambda=\lambda(\tau,\theta)$ and $d\Sigma$ denotes the surface 
 element in the space supported by the two Killing vectors. Polarized 
 Gowdy waves form a special case, which permit a reduction to
 \begin{eqnarray}
 d\Sigma^2=e^{-\tau}\left(e^Pd\sigma^2+e^{-P}d\delta^2\right).
 \label{EQN_G2}
 \end{eqnarray}
 Here $P$ satisfies a linear wave-equation
 $P_{\tau\tau}=e^{-2\tau}P_{\theta\theta}$; a long wave-length solution is
 \begin{eqnarray}
 P_0(\tau,\theta)=Y_0(e^{-\tau})\cos\theta,
 \label{EQN_G3}
 \end{eqnarray}
 where $Y_0$ is the Bessel function of the second kind of order zero.
 This leaves
 \begin{eqnarray}
 \lambda(\tau,\theta)=\frac{1}{2}Y_0(e^{-\tau})Y_1(e^{-\tau})
        e^{-\tau}\cos2\theta
       +\frac{1}{2}\int^1_{e^{-\tau}}\left(Y_0^{\prime 2}(s)+Y_0^2(s)\right)
	   sds.
\label{EQN_G4}
\end{eqnarray}
A spectrally accurate numerical integration is described in
\cite{mvp97}.

The implicit equation (\ref{EQN_B}) for the dynamical variables 
$(h^{n+1}_{ij},N_a^{n-1})$ has been implemented numerically. We have 
done so by solving for the all ten components 
$(h^{n+1}_{ij},N_a^{n-1}$)
using Newton iterations on these variables. This procedure uses a
numerical evaluation of the Jacobian
\begin{eqnarray}
J_{AB}=\frac{\partial R_{A}}{\partial U_B} 
\end{eqnarray}
where the capital indices $A,B=1,2,\cdots,10$ refer to the labeling
\begin{eqnarray}
\begin{array}{rl}
R_A& =(R_{11},R_{22},R_{33},R_{44},R_{12},R_{13},R_{14},R_{23},R_{24},R_{34}),\\
U_B& =(N_1^{n-1},h_{11}^{n+1},h_{22}^{n+1},h_{33}^{n+1},N_2^{n-1},N_3^{n-1},
       N_4^{n-1},h_{23}^{n+1},h_{24}^{n+1},h_{34}^{n+1}).
\end{array}
\end{eqnarray}
The Ricci tensor (\ref{EQN_R}) has been implemented by second-order
finite differencing, such that it remains quasi-linear in the second
derivatives. In particular, the Christoffel symbols
\begin{eqnarray}
\Gamma_{ab}^c = \frac{1}{2}g^{ce}(g_{cb,a}+g_{ac,b}-g_{ab,c})
\end{eqnarray}
is obtained by symmetric finite-differencing on the metric
components, and itself differentiated by the product rule following
individual numerical differentiations of $g^{ab}$ and 
$(g_{cb,a}+g_{ac,b}-g_{ab,c})$.
The choice of future gauge $N_a^{n+1}$ is provided by the 
the components
\begin{eqnarray}
g_{at}=(e^{(\tau-\lambda)/2},0,0.0)
\end{eqnarray}
of the analytical line-element 
(\ref{EQN_G1}-\ref{EQN_G4}), which facilitates error analysis
by direct comparison of the numerical results with the analytic 
solution. It will be appreciated
that in principle other choices of $N_a^{n+1}$ can be made. 

Fig. 1 shows numerical results for evolution of initial data
on the interval $0\le\tau\le 4$. The results show that {\em all} Einstein
equations in the form of $R_{ab}=0$ are satisfied with arbitrary
precision, while the metric components are solved accurately to 
within one percent. The asymptotic behavior of the implicit
corrections to the lapse functions are shown in Fig. 2. Note that
these corrections are finite to first-order in $\Delta t$: the
corrections $\delta N$ on the past lapse satisfy
\begin{eqnarray}
\frac{\delta N}{N}=\frac{N^{n+1}(\tau_n)-N^{n-1}(\tau_{n+2})}{N^n(\tau_{n+1})}=
O\left(\Delta\tau\right).
\end{eqnarray}
This first-order dependence is a testimonial to the fact that the
lapse function appears in the Einstein equations to first-order
in time.\\
\mbox{}\\
{\bf Theorem 2.} {\em A choice of gauge in the future and a dynamical gauge
in the past obtains a discretization of general relativity 
consistent with energy and momentum conservation, permitting all
ten Einstein equations to be solved in ten dynamical variables in the
presence on non-exact projections of the four-covariant metric on the
past hypersurface of constant time.}

\begin{figure}
\centerline{\epsfig{file=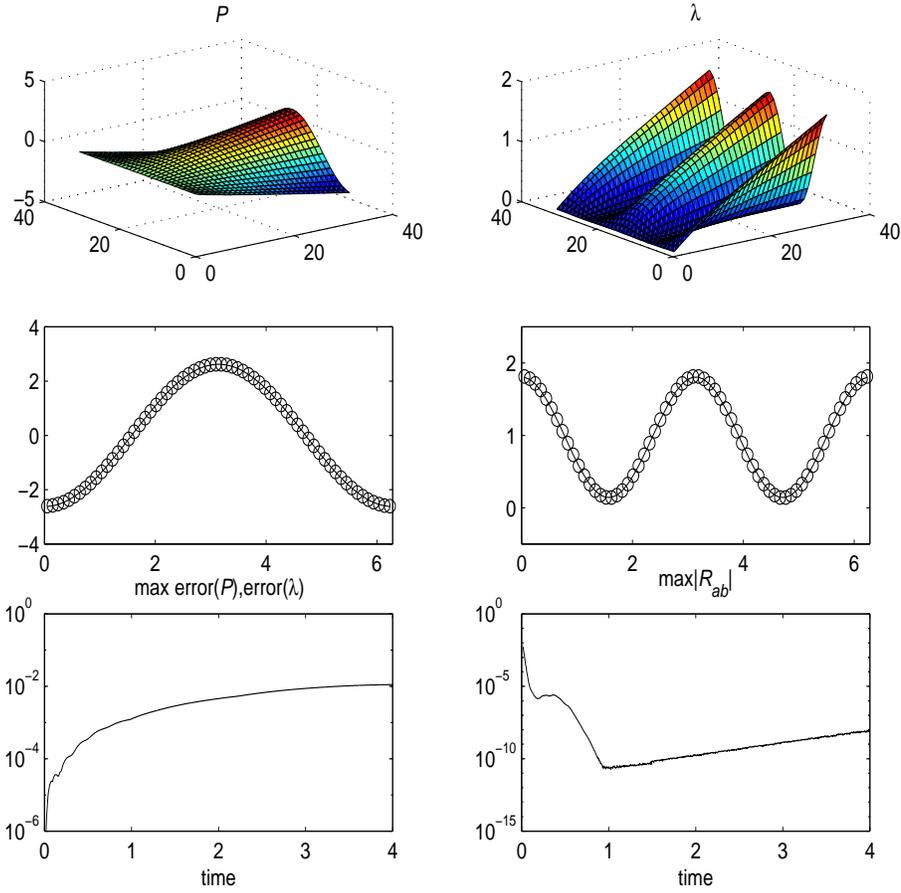,width=120mm,height=120mm}}
\caption{Shown is the simulation for $0\le\tau\le 4$ of
the polarized Gowdy wave. The solutions $P(\tau,\theta)$ and 
$\lambda(\tau,\theta)$ are displayed as a function of 
($\tau,\theta)$ (upper windows).
The middle windows display the solutions for $\tau=4$,
wherein the circles denote the numerical solution and the
solid lines the analytical solution.
The $\tau-$evolution of the errors (lower windows) 
are computed relative
to the analytical solution to Gowdy's reduced wave equation. 
The simulations use a discretization of $\theta$ by $m_1=64$ points and 
the $\tau-$interval by $m_2=1024$ time-steps. 
Particular to the proposed numerical algorithm is a dynamical gauge
in the past and a prescribed gauge in the future. This permits
satisfying {\em all} of the discretized Einstein equations 
$R_{ab}=0$ to within arbitrary precision by Newton iterations.
The slight increase in the error of about $10^{-10}$ reflects
the exponential growth of the analytic solution, because the
Gowdy cosmology evolves towards a singularity. 
(Reprinted from van Putten, M.H.P.M., {\em Class. Quantum Grav.}, 19, L51,
\copyright IOP 2002.)}
\end{figure}

\begin{figure}
\centerline{\epsfig{file=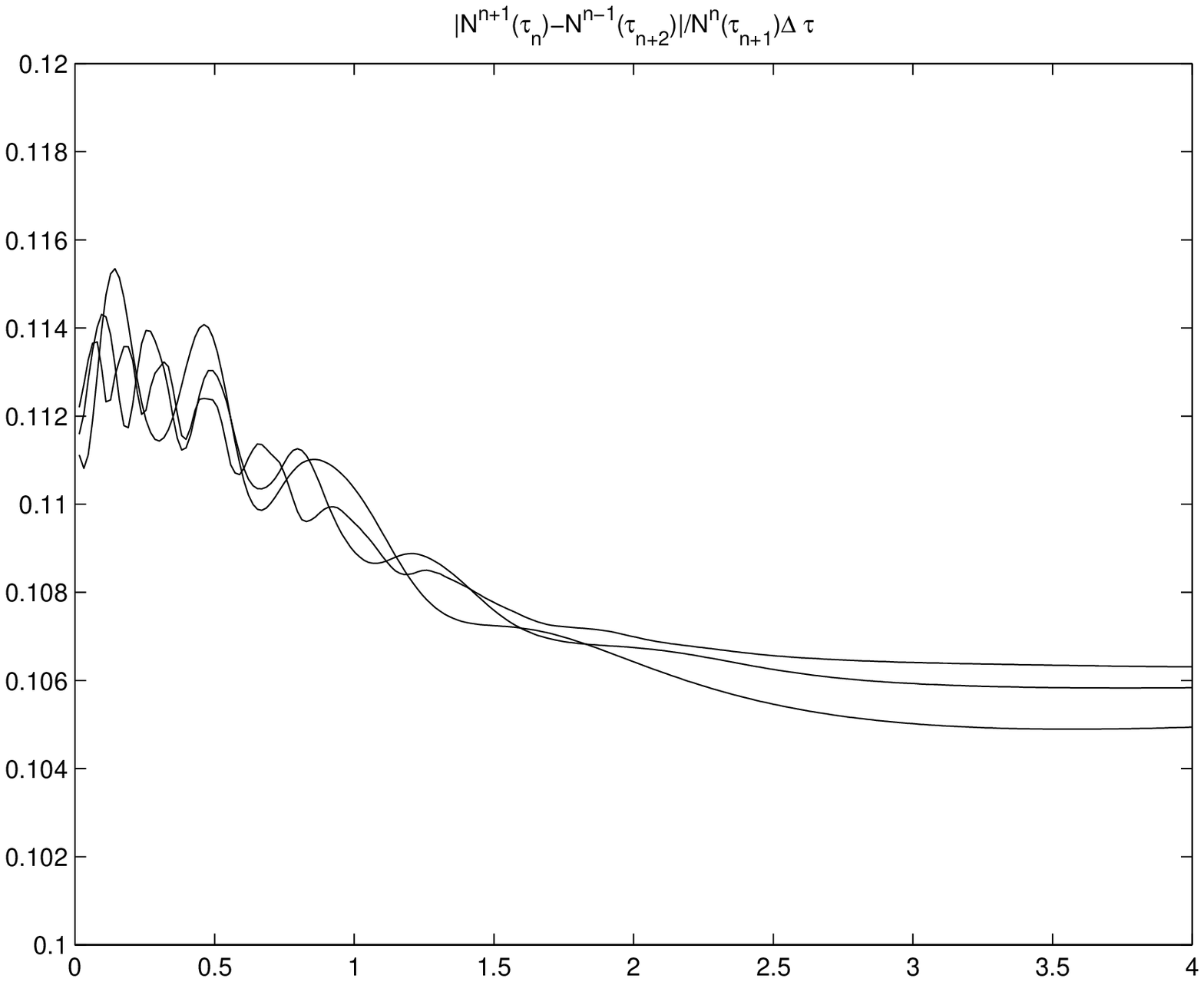,width=120mm,height=120mm}}
\caption{ Shown are the self-consistent corrections on the
slicing $t=t_{n+1}$, introduced by the difference between the past gauge 
$N^{n-1}(t_{n+2})$ to the hypersurface $t=t_{n+2}$ and the earlier future gauge 
$N^{n+1}(t_n)$ to the hypersurface $t=t_{n}$. The three curves refer
to different discretizations $m_1=16,32$ and $64$ points with, respectively,
$m_2=256,512$ and 1024 time-steps. These similar results for various
discretizations indicate asymptotic
behavior consistent with the first-order appearance of the lapse function 
in the Einstein equations. A first-order accuracy in lapse introduces a
commensurate offset in slicing or, equivalently, an offset in the coordinate $t$.
Similar results obtain for the same spatial discretizations $m_1=16,32$ and
$m_1=64$ with time-steps at one-half the value, i.e., using
$m_2=512, 1024$ and, respectively, $m_2=2048$ time-steps.
(Reprinted from van Putten, M.H.P.M., {\em Class. Quantum Grav.}, 19, L51,
\copyright IOP 2002.)}
\end{figure}

\section{Summary and conclusion}

Well-posed numerical relativity is a long-standing challenge in the
calculation of wave-forms for astrophysical sources of gravitational radiation. 
A necessary condition for stable numerical relativity is accurate conservation
of the energy and momentum constraints (``integration on thin ice"). This has 
been pursued by implementing these constraints dynamically \cite{bro99,shi00,sie01}. 
Here, we have explored a consistent discretization for exact conservation of 
energy-momentum constraints using a choice of gauge in the future and a
dynamical gauge in the past. This permits integration of {\em all} ten
Einstein equations, while allowing for in-exact projections 
of the four-covariant metric onto the surfaces of foliation of spacetime.
The simulation of a nonlinear one-dimensional Gowdy wave by 
implicit time-stepping according to the ten discretized vacuum Einstein 
equations (\ref{EQN_B}) serves to illustrate a numerical implementation. 

A major open problem is obtaining
sufficient conditions for stability. In this presented approach,
it becomes of interest to consider novel definitions of 
the future gauge as a function of present gauge. We leave this
as a suggested direction for future development.
In light of the recent discussion by Gambini and Pullin (2002) \cite{gam01},
the question arises: is well-posed numerical relativity
related to consistent discretization in quantum gravity?

Evolving eternal Schwarzschild black holes in 3+1 may serve as a
test problem for these developments \cite{kid01}.
More generally, it would be of interest to consider exact conservation
of energy-momentum in higher 
dimensions, perhaps using a combination of any of the modern 
hyperbolic formulations and efficient elliptic solvers.

{\bf Acknowledgments.} The author thanks the APCTP at Pohang
University for hosting a very stimulating Winter School on 
black hole astrophysics and stimulating discussions with L. Wen. 
This research is supported by NASA Grant 5-7012 and an MIT C.E. Reed 
Award. 
\mbox{}\\

\noindent {\large References}


\end{document}